\documentstyle[epsf,12pt]{article}
\begin{document}
\setcounter{page}{0}
\thispagestyle{empty}

\begin{center}
{\bf Imaginary part of the electromagnetic lepton form factors}

\vspace{15mm}

Leo.~V.~Avdeev and M.~Yu.~Kalmykov
\footnote{
E-mail: kalmykov@thsun1.jinr.dubna.su}

\vspace{15mm}

{\it
Bogoliubov Laboratory of Theoretical Physics, \\
Joint Institute for Nuclear Research, \\
Dubna, Moscow Region 141980, Russian Federation}

\vspace{5mm}

\end{center}

\begin{abstract}

The charge $F_1(0)$ and the magnetic $F_2(0)$ form factors of heavy
charged leptons have been shown in the framework of the perturbation
theory to have imaginary part. The imaginary parts of the form
factors for muon and tau lepton have been calculated at the two-loop
level in the Standard Model. The effects where these imaginary parts
could manifest themselves are discussed.

\end{abstract}

\vspace*{10pt}

{\it PACS number(s)}:
13.40.Em, 13.40.Gp, 13.10.+q, 13.35.-r, 12.15.Lk


\pagebreak
\section{Introduction}
\vspace*{-0.5pt}

   Recent high precision experiments to verify the Standard Model of
electroweak interactions require, on the side of the theory,
higher precision calculation of various physical quantities. The
charge and the magnetic form factors of the photon-lepton-lepton
vertex belong, in particular, to the class of such quantities. These
form factors are the fundamental quantities in the elementary
particles physics. Their importance derives from the fact that
these quantities can be measured very precisely, and at the same
time are calculable from first principles.

Let $\Gamma_\mu(p_1,p_2)$ be the vertex amplitude of the
photon-lepton-lepton process. If $u_1, \overline{u}_2$ are the
spinors describing the on-shell initial and final lepton states
with mass $m$, the most general form of the matrix element is

\begin{equation}
\overline{u}_2 \Gamma_\mu u_1
= -i e \overline{u}_2 \Bigl [
F_1(t) \gamma_\mu + \frac{i}{4m} F_2(t) \left(
\gamma_\mu \gamma_\nu - \gamma_\nu \gamma_\mu \right) q^\nu
+ \frac{1}{m} F_3(t)q^\nu
+ \cdots  \Bigr] u_1 ,
\end{equation}

\noindent
where $''\cdots''$ denote the terms proportional to $\gamma_5$;
$q^\mu = p_2^\mu - p_1^\mu$ is the photon momentum;
$t = q^2$; $p_1^2 = p_2^2 = -m^2$,
the space-time we are working in is the N-dimensional Euclidean one.
As can be verified by using the Gordon decomposition,
the t-dependent form factors can be interpreted in a standard way for
$t=0$ and with the lepton on the mass shell, at the same time, i.e.,
$F_1(0)$ is the electric charge of the lepton, $F_2(0)$ is the static
anomalous magnetic moment of the lepton. Using the method described
in Ref.  \cite{projector}, $F_1(0)$ and $F_2(0)$ can be directly
extracted from the vertex amplitude.  Expanding the amplitude
$\Gamma_\mu(p,q)$ up to the first order in $q$

$$
\Gamma_\mu(p,q) = \Gamma_\mu(p,0) + q^\nu \frac{\partial}{\partial
q^\nu}
\Gamma_\mu(p,q)|_{q=0} \equiv V_\mu(p) + q^\nu T_{\mu \nu} (p),
$$

\noindent
we have:

\begin{eqnarray}
F_1(0) & = & \frac{-1}{4m^2 e}
Sp \Bigl[ \left( i \hat p - m \right) p_\mu V_\mu(p) \Bigr] ,
\nonumber \\
F_2(0) & = &
 -  \frac{1}{8m \left(N-2 \right) \left(N-1 \right) e}
Sp \Bigl[
\left( i \hat p - m \right)
\left( \gamma_\mu \gamma_\nu - \gamma_\nu \gamma_\mu \right)
\left( i \hat p - m \right) T_{\mu \nu} \Bigr]
\nonumber \\*
&& + \frac{i}{4m^2 \left( N-1 \right) e}
Sp \Bigl[ \left( m^2 \gamma_\mu + N \hat p p_\mu
+ im \left( N-1 \right)  p_\mu  \right) V_\mu(p) \Bigr] .
\nonumber
\end{eqnarray}

\noindent
Therefore, the calculation of the charge and magnetic form
factors of the lepton reduces, after differentiation and contractions
with projection operators, to diagrams of the propagator type with
external momentum on the lepton mass shell.  All the $F_i(0)$ must be
real thanks to hermicity of the electromagnetic current. However,
this statement is fully correct only for QED. In the framework of
the Standard Model all the $F_i(0)$ for unstable leptons have
imaginary parts. This paper is aimed at calculation of the
imaginary parts of the corresponding form factors, and at discussion
of their identification in real experiments.

Our plan is the following: In Sect.2 we consider the toy model as
an example and show that the instability of fermions leads to the
imaginary parts of the electromagnetic form factors. Sect.3
represents the calculation of the imaginary parts of the
electromagnetic form factors of leptons at the two-loop level
in the Standard Model. Sect.4 is devoted to discussion of their
physical meaning.

\section{The toy model}

Let us consider a ``toy'' model where heavy and massless charged
spinors ($\Psi$ and E, respectively), the photon $A_\mu$, and a light
neutral scalar field $\Phi$ are included.  The scalar has the Yukawa
coupling $\left( y \right)$ to the heavy and massless spinors.  The
Lagrangian of this model can be written (in the Euclidean space-time)
as

\begin{eqnarray}
{\it L} & = & \frac{1}{2} \partial_\mu \Phi \partial^\mu \Phi
+ \frac{1}{2} m^2 \Phi^2
+ \frac{1}{4} \left(\partial_\mu A_\nu - \partial_\nu A_\mu \right)^2
+ \frac{1}{2 \xi} \left( \partial_\mu A^\mu \right)^2
\nonumber \\*
& + & \bar{\Psi} \left( \hat{\partial} + M \right) \Psi
+ \bar{E} \hat{\partial} E
+ i e \left( \bar{\Psi} \hat A \Psi + \bar{E} \hat A E \right)
+ y \Phi \left( \bar{\Psi} E + \bar{E} \Psi \right),
\label{toy-model}
\end{eqnarray}

\noindent
where $e$ is the electric charge; $\xi$ is the gauge parameter.
In this model the leptons' instability displays itself in the
electromagnetic form factors even at the one-loop level --
which significantly facilitates the calculation.

\begin{figure}[bth]
\centerline{\vbox{\epsfysize=30mm \epsfbox{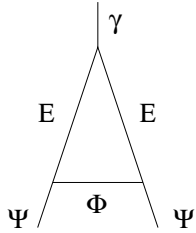}}}
\caption{\label{toy1} The Yukawa contribution to the electromagnetic
form factors of the heavy spinor within the toy model.}
\end{figure}

The contribution of the diagram given in Fig.\ref{toy1} to the
electromagnetic form factors of the heavy spinor can be easily
found:

\begin{eqnarray}
F_1(0) & = & \frac{y^2}{32 \pi^2} \Biggl\{ \frac{1}{\varepsilon}
- \ln \left(\frac{m^2}{\mu^2} \right)
- \ln \left(1 - \frac{M^2}{m^2} \right)
\nonumber \\*
& + &
\frac{m^2}{M^2} \left[ 3 - 2 \ln \left(1 - \frac{M^2}{m^2} \right)
\right]
+ 3 \left( \frac{m^2}{M^2} \right)^2 \ln \left(1 -
\frac{M^2}{m^2} \right) \Biggr \},
\nonumber  \\
F_2(0) & = & - \frac{y^2}{16 \pi^2} \left[ 1 + 2 \frac{m^2}{M^2}
+ 2 \left( \frac{m^2}{M^2} \right)^2 \ln \left(1 - \frac{M^2}{m^2}
\right) \right] ,
\end{eqnarray}

\noindent
so that

\begin{eqnarray}
{\it Im} F_1(0) & = & -i \frac{y^2}{32 \pi} \left[
1 +  2 \frac{m^2}{M^2} - 3 \left( \frac{m^2}{M^2} \right)^2
\right] ,
\label{toyF1}  \\
{\it Im }F_2(0) & = & - i \frac{y^2}{8 \pi} \left( \frac{m^2}{M^2}
\right)^2 ,
\label{toyF2}
\end{eqnarray}

\noindent
where the sign of the imaginary part is defined by the
"causal" $i0$-prescription, $\ln(-m^2)=\ln(m^2)+i \pi$. Since we are
interested only in the imaginary parts of the form factors, the
diagram with the virtual photon is omitted. It is obviously why
the imaginary parts of the form factors may appear. The vertex diagram
reduces, after differentiation and contractions with projection
operators, to the sum over the propagator type diagrams with external
momentum on the mass shell of the heavy particle. The last type of
diagrams generates imaginary part proportional to the decay
width of the heavy particle.  The relations between the imaginary
part of the form factors and the corresponding decay width of the
heavy fermion may be naturally suggested to exist and to be simple

\begin{equation}
{\it Im} F_i(0) = C_i \Gamma,
\label{trivial}
\end{equation}

\noindent
where $\{ C_i \}$ are the constants. This relation can be easily
checked. Let us consider the one-loop-particle irreducible two-point
function $\Sigma(\hat{p},m)$ of the heavy fermion with the massless
fermion and the light neutral scalar particle inside the loop.
The decay width $\Gamma$ of the particle is proportional to the
imaginary part of the propagator at the point $p^2 = -M_P^2$,
where $M_P$ is the pole mass of the particle. Solving the equation
$i \hat{p} + m - \Sigma (\hat{p},m)$, we obtain

\begin{eqnarray}
\Gamma \sim {\it Im} \Sigma (M_P, M_P) = i \frac{y^2}{32 \pi}
\left( 1 - \frac{m^2}{M^2} \right)^2.
\label{G}
\end{eqnarray}

\noindent
The comparison of the Eqs.(\ref{toyF1}), (\ref{toyF2}), and (\ref{G})
shows that the relation (\ref{trivial}) does not work in the
general case. It is natural to expect that the imaginary parts of the
corresponding form factors are related to the decay width in a more
complicated nonlinear way.

An additional contribution into the form factors $F_i(t)$ like $t \ln (-t)$
appears at small non-zero $q^2=t$. In the space-time region $t>0$
that contribution results in imaginary part due to zero threshold
with respect to t (see Fig.\ref{toy1}). However, this value is
proportional to $i \pi t$, that is why it disappears at $t=0$.

\section{The imaginary part of the lepton's form factors}

\begin{figure}[bth]
\centerline{\vbox{\epsfysize=100mm \epsfbox{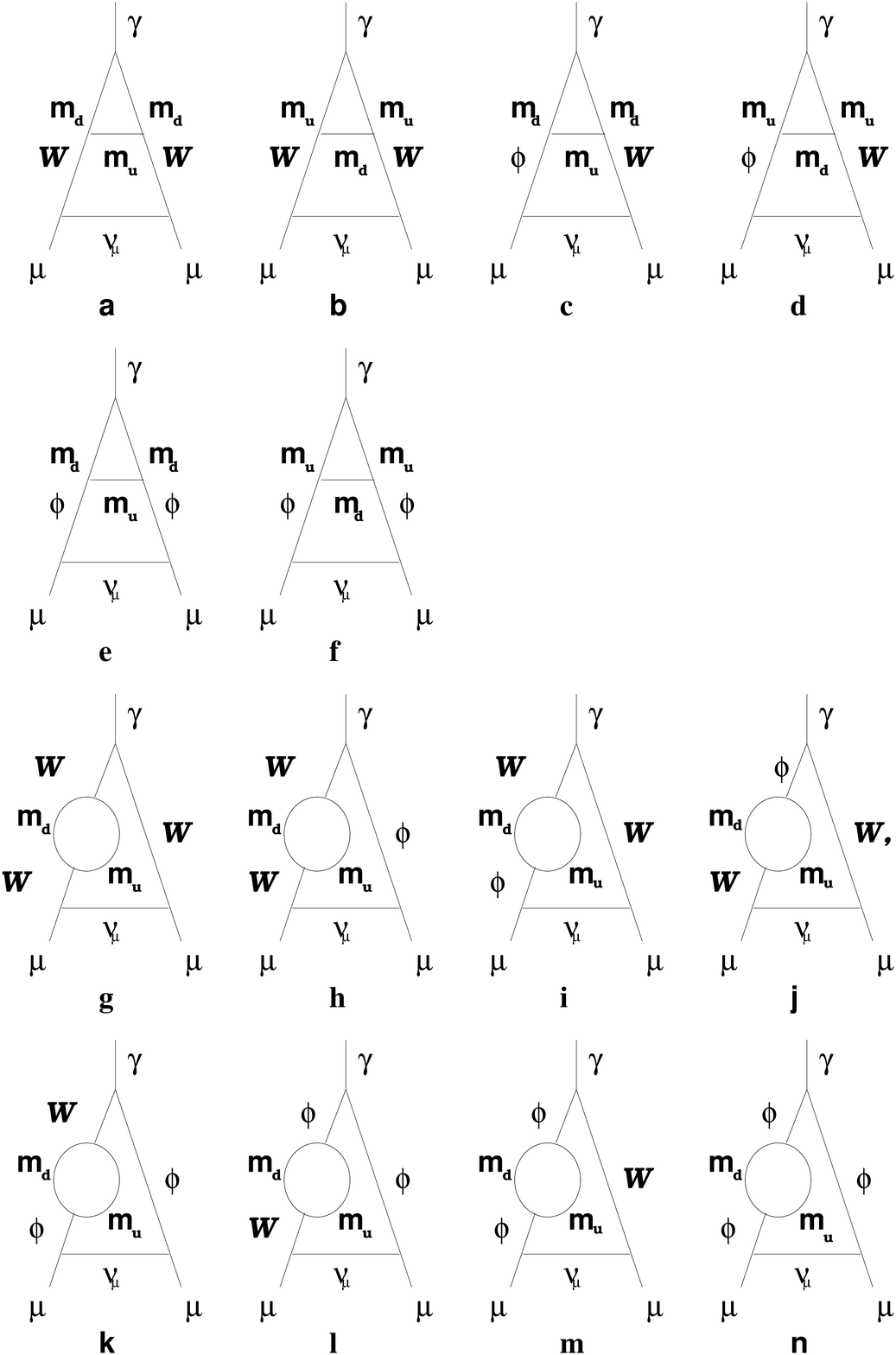}}}
\caption{\label{im1} The two-loop diagrams contributing to the
imaginary part of the electromagnetic form factors of a heavy
lepton within the Standard Model.}
\end{figure}

\begin{figure}[bth]
\centerline{\vbox{\epsfysize=40mm \epsfbox{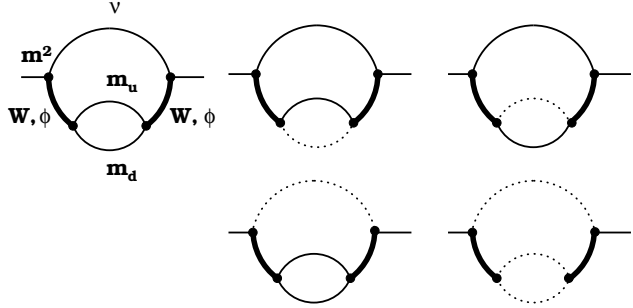}}}
\caption{\label{im2} The structure of large-mass expansion. Bold and
thin lines correspond to heavy-mass and light-mass (massless)
propagators, respectively. Dashed lines indicate the lines omitted in
the original graph to yield the subgraph.}
\end{figure}

\begin{figure}[bth]
\centerline{\vbox{\epsfysize=60mm \epsfbox{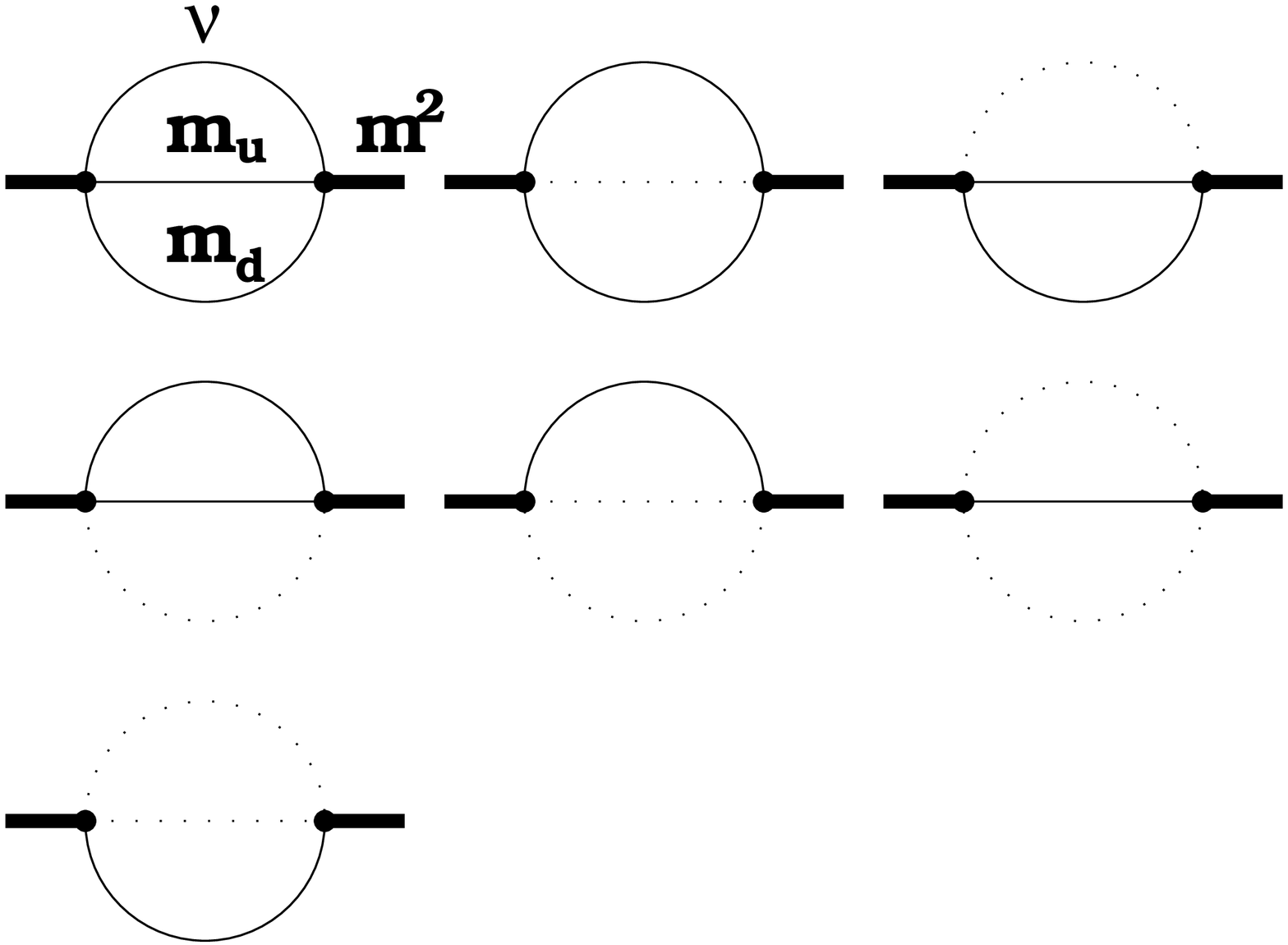}}}
\caption{\label{im3} The structure of large-momentum expansion.
The notations are the same as in Fig.\ref{im2}.}
\end{figure}

Now we concentrate on the charge and the magnetic form factors of
leptons in the framework of the Standard Model.
The imaginary parts of given form factors arise only at the two-loop
level. If we restrict ourselves to the perturbation theory only, the
light quarks should be considered as fermions with the corresponding
masses. The corresponding diagrams are presented in Fig.  \ref{im1}.
Since we work in the Feynman gauge, the would-be-Goldstone $\phi$ has
the same mass $M_W$ as the charge boson $W$. All the rest dimensional
parameters in these diagrams are small in compared with $M_W$.
So, the asymptotic expansion method \cite{asymptotic} is wholly
suitable for calculation of the diagrams under consideration.
The rules of purely Euclidean asymptotic expansions are as follows.
The expansion is a sum over 'ultraviolet' subgraphs of the diagram.
An ultraviolet subgraph must contain all lines with large masses, the
points where large external momenta (if any) flow in/or out. The large
momenta ought to go only through the ultraviolet subgraph, and obey
the momentum conservation law.  And last, an ultraviolet subgraph
should be one-particle irreducible with respect to lines with small
and zero masses, although may consist of several disconnected parts.
An ultraviolet subgraph is Taylor-expanded in its small parameters
(external momenta and internal masses), and then shrunk to a point and
inserted in the numerator of the remaining Feynman integral.
The two consecutive expansions are carried out to calculate the
diagrams given in Fig. \ref{im1}. The first of them is the large-mass
expansion with respect to the heavy mass $M_W$. The set of the
corresponding subgraphs is shown in Fig. \ref{im2}. This expansion
says that only the last subgraph consists of an imaginary part.
The large-mass expansion leads to

\begin{eqnarray}
F & = & \sum_{l=0}^\infty \left( \frac{1}{M_W^2} \right)^l
\sum_{k=0}^2 \Phi_{l,k}(m_u^2, m_d^2, m^2)
\ln^k \frac{M_W^2}{\mu^2} ,
\nonumber
\end{eqnarray}

\noindent
where $F$ is the initial Feynman integral, $m_u^2, m_d^2$  are
the fermion masses in the loop, and $m^2$ is the mass of an external
lepton. $\Phi_{l,k}(m_u^2, m_d^2, m^2)$ are some complicated
functions of their arguments in the general case. The maximum power
of logarithm is defined by the highest degree of divergences
(ultraviolet, infrared, collinear) in the subgraphs (equals 2 in
our cases); $\mu^2$ is the subtraction point.  As a result of the
asymptotic expansion with respect to the heavy mass, the two-loop
bubble integrals and the propagator-type integral with the different
masses $(m_u^2, m_d^2)$, with the external momentum $(p^2 = -m^2)$ and
with the reduced number of internal lines arise. The second type of
the integrals mentioned can be calculated by using the asymptotic
expansion with respect to the large external momentum, which is
the particular case of Euclidean expansion.  The structure of the
asymptotic expansion in this case is given in  Fig.  \ref{im3},
so that

\begin{eqnarray}
\Phi_{l,k}(m_1^2, m_2^2, m^2) & = &
\left( m^2 \right)^l
\sum_{a,b=0}^\infty
\left( \frac{m_1^2}{m^2} \right)^a
\left( \frac{m_2^2}{m^2} \right)^b
\nonumber \\*
&&
\sum_{p=0}^{2-k}
\left(
A_p^{~ab} \ln^{p} \frac{m_1^2}{\mu^2}
+ B_p^{~ab} \ln^{p} \frac{m_2^2}{\mu^2}
+ C_p^{~ab} \ln^{p} \frac{m^2}{\mu^2}
\right) ,
\nonumber
\end{eqnarray}

\noindent
where $\{A_k^{~ab}, B_k^{~ab}, C_k^{~ab} \}$ are the numbers.
All the calculations are performed by means of the package TLAMM
\cite{TLAMM}. Since the imaginary part of the corresponding form
factors does not include divergences, no additional renormalization
is required.

\section{Discussion and conclusions}

The imaginary parts of the charge and the magnetic form factors of
leptons in the leading order of the Standard Model have the
following form:

\begin{eqnarray}
{\it Im} F_1(0) & = &
i \frac{G_F^2 m^4}{ 8 \pi^3} N_c
\sum_f
\Biggl [
\left( Q_d - Q_u \right)
\biggl( \frac{5}{48} - \frac{r_1^2 + r_2^2}{2}
+ O(r_k^4 \ln r_k) \biggr)
\nonumber \\
&&
+ \frac{m^2}{M_W^2}
\biggl \{
\left( Q_d - Q_u \right)
\left(
\frac{1}{30} - \frac{r_1^2 + r_2^2}{6}
\right)
- \frac{13}{240}
+ \frac{17}{48} \left( r_1^2 + r_2^2 \right)
\nonumber \\
&&
+ O(r_k^4 \ln r_k)  \biggr \} + O(m^4/M_W^4) \Biggr ],
\nonumber \\
{\it Im} F_2(0) & = & i \frac{G_F^2 m^4}{ 8 \pi^3} N_c
\sum_f
\Biggl [
- \frac{Q_d + 3 Q_u}{12}
+ \frac{r_1^2 + r_2^2}{3}
\left( Q_d + 2 Q_u \right)
+ O(r_k^4 \ln r_k)
\nonumber \\
&&
+ \frac{m^2}{M_W^2}
\biggl \{
\frac{Q_d - Q_u}{40}
- \frac{1}{40} + \frac{r_1^2 + r_2^2}{4}
+ Q_d
\left(
-\frac{5}{36} r_1^2 + \frac{r_2^2}{2}  - \frac{r_1^2}{3}
\ln r_1
\right)
\nonumber \\
&&
+ Q_u \left(
\frac{r_1^2}{6} - \frac{17}{12} r_2^2  - r_2^2 \ln r_2
\right)
+ O(r_k^4 \ln r_k) \biggr \} + O(m^4/M_W^4) \Biggr ],
\label{leading}
\end{eqnarray}

\noindent
where all the fermions with $T_3 = 1/2$ are called as u-fermions
with the electric charge $Q_u = 2/3$ (in units of the positron
charge) and with the mass $m_u$. Correspondingly,
the fermions with $T_3 = -1/2$  are d-fermions with the charge
$Q_d = -1/3$ and the mass $m_d$; m is the mass of external lepton;
$\sum_f$ is the sum over all the fermions with $m_f^2 < m^2$;
$r_1 = m_d/m$ and $r_2 = m_u/m$.
To extract the contribution of lepton with the mass $m_l$, we
should accept that $m_u = 0, Q_u = 0, Q_d = -1, m_d = m_l$ ;
$N_c$ is the color factor which is equal to 3 for quarks and 1
for leptons, respectively; $G_F/\sqrt{2} = g^2/8/M_W^2$, and
the corresponding Cabibbo-Kobayashi-Maskawa matrix is equal to
{\bf I}. When considering light quarks as internal fermions it should
be pointed out that the perturbative quantum field theory leads to
incorrect results.  In particular, the muon has no hadronic mode of
decay. So, at evaluation of the imaginary part of the muon
electromagnetic form factors we take into account only the lepton
contribution:

\begin{equation}
{\it Im} F_1^\mu(0) \sim
- i \frac{5}{384} \frac{G_F^2 m^4_\mu}{\pi^3},
~~~~~
{\it Im} F_2^\mu(0) \sim i \frac{G_F^2 m^4_\mu}{96 \pi^3}.
\label{muon}
\end{equation}

\noindent
The tau-lepton is the only presently known lepton massive enough
to decay into hadrons \cite{pich}. Taking the decay mode
$\tau^-  \rightarrow \nu_\tau d \bar{u}$ into account, we obtain
the following evaluation for the imaginary parts of the corresponding
form factors of the tau-lepton:

\begin{equation}
{\it Im} F_1^\tau(0) \geq
- i \frac{25}{384} \frac{G_F^2 m^4_\tau}{\pi^3},
~~~~~
{\it Im} F_2^\tau(0) \geq - i \frac{G_F^2 m^4_\tau}{24 \pi^3}.
\label{tau}
\end{equation}

\noindent
The leading, over the mass, contribution into imaginary parts of
electromagnetic form factors can be found in the Fermi theory of
electromagnetic interaction. Since we are interested in the renormalization
of the corresponding form factors in the on-shell scheme, we choose
the renormalizable theory GWS for the calculation.

\begin{figure}[bth]
\centerline{\vbox{\epsfysize=40mm \epsfbox{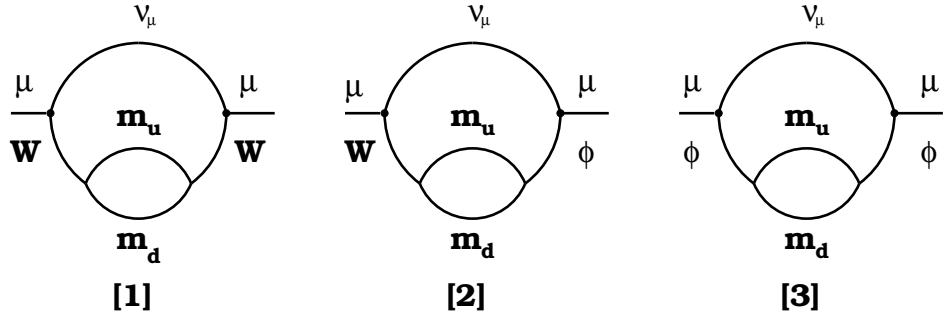}}}
\caption{\label{prop1} The diagrams contributing to the imaginary part
of the lepton's wave-function renormalization constant in the
Standard Model.}
\end{figure}

It is worthy to discuss the physical meaning of the imaginary parts of
the corresponding form factors. For this purpose let us first consider
the charge form factor. In QED the electric charge of electron is usually
fixed in the Thompson limit of Compton scattering. This definition ensures
that {\bf \it e} indeed represents the electric charge which describes
the interaction between electron and electromagnetic fields in the
classic electrodynamics. In the on-shell renormalization scheme
\cite{on-shell} the charge of electron is formally defined as the coupling
of a photon with zero momentum to an on-shell lepton.
This statement is commonly assumed to be valid also in the Standard Model,
and has been recently generalized to the case of arbitrary charged particle
\cite{compton}. There are reasons to suppose that this statement works also
for any charged lepton at arbitrary order of the perturbation theory.
This condition is usually written in the on-shell scheme as follows:

\begin{equation}
F_1^{\it ren}(0) = 1,
\label{on-shell1}
\end{equation}

\noindent
where
$$\Gamma_{\mu}^{\it ren} =
= -i e \Bigl [
F_1^{\it ren}(t) \gamma_\mu + \frac{i}{4m} F_2^{\it ren}(t) \left(
\gamma_\mu \gamma_\nu - \gamma_\nu \gamma_\mu \right) q^\nu \Bigr]
= \sqrt{Z_{\gamma}} \gamma_0 \sqrt{Z_f} \gamma_0 \Gamma_{\mu}
\sqrt{Z_f}.
$$

\noindent
Here $Z_f$ and $Z_\gamma$ are the renormalization constants of the
on-shell lepton and photon wave functions, respectively;
$\Gamma_{\mu}$ is the bare vertex photon-lepton-lepton
where $\gamma-Z$ mixing is taken into account. At the two-loop
level the wave function renormalization constant $2 Z_f = Z_R + Z_L +
\gamma_5 (Z_L - Z_R)$ is the complex value. The corresponding finite
imaginary part arises from the diagrams depicted in Fig. \ref{prop1}.
As was shown in Refs. \cite{im}, the only hermitian (real) part of
the wave function renormalization constant contributes to physical
observables. The antihermitian (imaginary) part is related to
transformation of the matrices of fermion fields from the weak
interaction eigenstates to the mass eigenstates. The renormalization
conditions, which fix these rotation matrices, are independent of all the
other on-shell renormalization conditions. It is noteworthy to say
also that  physical observables cannot be constructed from these
matrices in the lepton sector of the Standard Model (in the quark
sector, the Cabibbo-Kobayashi-Maskawa matrix is such an observable).
Therefore, the finite antihermitian (imaginary) part of the lepton
wave function renormalization constant can be omitted from the
renormalization of the charge and the magnetic form factors. So, at
the two-loop level the only $F_1(0)$ gives an imaginary contribution in
$\Gamma_{\it ren}$.  The definition (\ref{on-shell1}) allows
one to find the relation between the bare charge of a fermion and the
physical charge {\it e}. In the Standard Model at the one-loop level,
this relation does not depend on the on-shell lepton --
which in turn means "charge universality"  \cite{charge}.
We understand "charge universality"  as the fact that
the relation between the bare charge and the physical one is
independent of the on-shell lepton. The presence of imaginary part
in the electromagnetic form factor $F_1^{\it ren}(0)$ leads to
the breakdown of "charge universality". Certainly, this problem
might be related with inconsistency of treatment of instable particles
in the standard perturbative quantum field theory \cite{veltman}.
In the framework of such a self-consistent theory (if it will be created),
the $F_1^{\it ren}(0)$ of a heavy lepton is expected not to consist of
imaginary part at all, and the "charge universality" principle will be
restored. Although there is still no such a self-consistent
theory, we should find a way to treat with imaginary parts arisen.
Since the amplitude for fermion scattering from an electric field is
proportional to the $F_1^{\it ren}(0)$, it is naturally to use
the following condition:

\begin{equation}
\left|F_1^{\it ren}(0) \right| = 1,
\label{new1}
\end{equation}

\noindent
instead of (\ref{on-shell1}). If the renormalization condition
(\ref{new1}) is taken into account, it is then easy to show that the
two-loop imaginary part is equivalent to the finite four-loop
correction into the relation between the bare and the physical charge
of a lepton. Therefore, the "charge universality" saves, up to the
four-loop level.

Now, let us concentrate on the magnetic form factor. It is commonly
assumed that $\frac{g-2}{2} = F_2(0)$. However, this statement is
correct only for stable particles. So, the anomalous magnetic moment
of instable particle is not defined good enough, both
from theoretical point of view and experimental one (it is difficult to
measure this value due to the finite life time of such a particle).
When analyzing a fermion scattering from a static vector field we obtain
that the matrix element is proportional to the sum $F_1(0)+F_2(0)$.
So, we should suppose that the following relation can be used for the
numeric estimation of the magnetic moment for unstable particle:

\begin{equation}
\left|F_1(0) + F_2(0) \right| = \frac{g}{2},
\label{new2}
\end{equation}

\noindent
where $g$ is the Land\' e-factor.
If Eq. (\ref{new1}) is taken into consideration, we obtain that the
imaginary parts of the charge and the magnetic form factors are
equivalent to the four-loop contribution to the anomalous magnetic
moment of lepton, and can be omitted in the two-loop results
\cite{amm}.

The presence of imaginary parts in electromagnetic form factors of
heavy leptons is a subject of only theoretical studies
at the present time, because their influence on electroweak processes
is extremely small \cite{decay}. In fact, their effect is equivalent
to four-loop radiative corrections -- which is beyond the modern
experimental precision.

In the Standard Model it is commonly believed that the charged
leptons are identical in all respect, excepting for their masses and
their distinct and conserved lepton numbers. This statement turns out
to be incorrect at the two-loop level where the instability of heavy
leptons results in the imaginary parts of the electromagnetic form
factors. The problem of how the particles instability should be
correctly taken into account at calculation of physical quantities in
the Standard Model is still open \cite{unstable}. Therefore,
additional relations between the arisen imaginary parts and physical
observables should be found (see, for example, \cite{sin}). This
paper deals with the calculation of the imaginary parts of the
two-loop electromagnetic form factors of charged leptons in the
Standard Model. The leading part of such contributions comes from
the diagrams with W-exchange. The contribution of the rest diagrams
is generated by additional powers of $(m/M_W)^2$ in accordance
with the Feynman's rules (note that imaginary part enters all the diagrams
in \ref{im1}).  As was demonstrated,
the imaginary parts of the form factors $F_i(0)$ are related with the
leptons instability, and there are no trivial relations between the
corresponding imaginary parts and the decay width.  To take the
imaginary parts of the corresponding form factors into account, the
conditions (\ref{new1}) and (\ref{new2}) have been suggested as a
generalization of the standard relations between the electric
(magnetic) form factors and the electric charge (the anomalous
magnetic moment) of a lepton, respectively.

\noindent
{\bf Acknowledgment}\\
\noindent
A short time before this paper was finished, to common grief,
my co-author Leo.~V. Avdeev dead.
We want to thank D.~Broadhurst, S.~Dittmaier, J.~Fleischer,
F.~Jegerlehner, E.~Kuraev, O.~Tarasov,
and O.~Veretin for fruitful discussions.
We are indebted to the referee for the useful comments which allowed us
to improve significantly this paper.
This work is partially supported by RFBR grant \# 98-02-16923
and by INTAS-2058.
M.~Kalmykov is grateful to the Physics Department of
the Bielefeld University for its warm hospitality where this paper has
been partly done.


\begin{thebibliography}{99}

\bibitem{projector}
R.~Z.~Roskies, E.~Remiddi and M.~J.~Levine
in {\it Quantum Electrodynamics} p. 163,
ed. T.~Kinoshita (World Scientific, Singapore, 1990).

\bibitem{asymptotic}
F.~V.~Tkachov,
INR preprints, P-0332, Moscow (1983);
P-0358, Moscow (1984);
S. G. Gorishny,
Preprint JINR E2-86-176, E2-86-177 (Dubna, 1986);
K.~G.~Chetyrkin,
Teor. Math. Phys. {\bf 75} (1988) 26;
ibid {\bf 76} (1988) 207;
MPI preprint, MPI-PAE/PTh-13/91, Munich (1991);
V.~A.~Smirnov,
Comm. Math. Phys. {\bf 134} (1990) 109;
{\it Renormalization and asymptotic expansions}
(Bikrh\"auser, Basel, 1991).

\bibitem{TLAMM}
L.~V.~Avdeev, J.~Fleischer,  M.~Yu.~Kalmykov and M.~N.~Tentyukov,
"Nuclear Instruments and Methods in Physics Research",
A {\bf 389} (1997) 343; Comp. Phys. Comm. {\bf 107} (1997) 155.

\bibitem{on-shell}
D.~A.~Ross and J.~C.~Taylor, Nucl. Phys. {\bf B51} (1973) 125;
G.~Passarino and M.~Veltman, Nucl. Phys. {\bf B160} (1979) 151;
A.~Sirlin, Phys. Rev. {\bf D22} (1980) 971;
J.~Fleischer and F.~Jegerlehner, Phys. Rev. {\bf D23} (1981) 2001;
S.~Sakakibara, Phys. Rev. {\bf D24} (1981) 1149;
K.~Aoki, Z.~Hioki, R.~Kawabe, M.~Konuma and T.~Muta,
Suppl. Progr. Theor. Phys. {\bf 73} (1982) 1.

\bibitem{compton}
S.~Dittmaier, Phys. Lett. B {\bf 409} (1997) 509.

\bibitem{pich}
A.~Pich, hep-ph/9704453.

\bibitem{im}
A.~Denner and T.~Sack,  Nucl. Phys. {\bf B347} (1990) 203;
B.~A.~Kniehl and A.~Pilaftsis, Nucl. Phys. {\bf B474} (1996) 286.

\bibitem{charge}
See the last Ref. in \cite{on-shell}.

\bibitem{veltman}
M. ~Veltman, Physica {\bf 29} (1963) 186.

\bibitem{amm}
T.~V.~Kukhto, E.~A.~Kuraev, A.~Schiller and Z.~K.~Silagadze,
Nucl. Phys. B{\bf 371} (1992) 567;
S.~Peris, M.~Perrottet and E.~de~Rafael,
Phys. Lett. B{\bf 355} (1995) 523;
A.~Czarnecki, B.~Krause and W.~Marciano,
Phys. Rev. D{\bf 52} (1995) R2619;
Phys. Rev. Lett. {\bf 76} (1996) 3267.

\bibitem{decay}
D.~J.~Silverman and G.~L.~Shaw, Phys. Rev. D{\bf 27} (1983) 1196;
J.~A.~Grifols and A.~M{\'e}ndez, Phys. Lett. B{\bf 255} (1991) 611;
J.~Biebel and T.~Riemann, Z.Phys. C{\bf 76} (1997) 53;
S.~S.~Gau, T.~Paul, J.~Swain and L.~Taylor, hep-ph/9712360.

\bibitem{unstable}
F.~V.~Tkachov,  hep-ph/9802307.

\bibitem{sin}
P.~Gambino and A.~Sirlin, Phys. Rev. D {\bf 49} (1994) R1160.

\end{thebibliography}
\end{document}